\documentclass[aps,prl,reprint,superscriptaddress]{revtex4-2}

\usepackage{graphicx}

\begin{document}


\title{Chirality enhancement using topology-designed 3D nanophotonic antennas}



\author{Atsushi Taguchi}
\email[]{taguchi@es.hokudai.ac.jp}
\affiliation{Research Institute for Electronic Science, Hokkaido University, Sapporo, Hokkaido 001-0020, Japan}
\author{Yamato Fukui}
\affiliation{Graduate School of Information Science and Technology, Hokkaido University, Sapporo 060-0814, Japan}
\author{Keiji Sasaki}
\affiliation{Research Institute for Electronic Science, Hokkaido University, Sapporo, Hokkaido 001-0020, Japan}


\date{\today}

\begin{abstract}
We explore chiroptical phenomena in 3D chiral nano-gap antennas using topology optimization. The characteristic helical geometries of the topology-designed antennas exhibit giant chiral dissymmetry ($g=-1.70$) considering the gap intensity, circular-to-linear polarization conversion, and circularly polarized light emission from a linear dipole coupled with the antenna. We observed that the spin angular momentum of light, flowing into the nanogap with opposite signs, locally amplifies optical chirality. These findings carry profound implications for the nanoscale control of complex light-matter interactions with structured light.
\end{abstract}


\maketitle

{\it Introduction.}---An object is chiral when it cannot be superimposed onto its mirror image. Light exhibits chirality, as illustrated by phenomena such as circularly polarized light (CPL) and optical vortices. The difference (or dissymmetry) in the interaction of left- and right-handed CPLs ({\it l}- and {\it r}-CPL) with a chiral molecule is often weak because molecules are much smaller than the wavelength of light. Nevertheless, chiroptical light--matter interactions play a crucial role in multiple fields, such as circular dichroism (CD) spectroscopy \cite{Barron.2004}, chiral synthesis and crystallization in chemistry \cite{Sugiyama.2020,Inoue.1992}, optical trapping \cite{Dholakia.2016,Sasaki.2021}, and spin state manipulation in quantum information technology \cite{Finley.2004,Awschalom.2006,Wu.2023}. Extensive efforts have been devoted to enhancing the chiroptical effect. This is exemplified by the utilization of chiral metallic nanostructures \cite{Kadodwala.2010,Okamoto.2013,Liu.2017,Giessen.2012txc,Giessen.2014mo8} and metamaterials \cite{Kuwata-Gonokami.2020xqx6}. However, the design of these structures has traditionally relied on heuristic approaches, largely dictated by a finite set of preassigned design parameters determined by humans. Considering the inherent complexity of chiral structured light, we delved into crafting a high-performing chiral nanophotonic structure.

Topology optimization (TO), initially proposed for mechanical designs \cite{Kikuchi.1988}, uses gradient-based searches rooted in Maxwell's equations \cite{Sigmund.201110j,Vuckovic.2013rxd}. TO enables the handling of large-scale structure designs, comprising hundreds to potentially billions of permittivity values at each spatial point, offering a considerable degree of design freedom \cite{Sigmund.2017}. This allows for designing structures by passing through topologically discontinuous deformation, which is not possible using traditional design strategies that only optimize dimensional design parameters and the outer shape. Notably, TO has enhanced the design of photonic waveguides \cite{Vuckovic.2015}, nanogap cavities \cite{Stobbe.2022}, and metallic structures for Raman scattering \cite{Johnson.2020}. Nonetheless, despite its remarkable flexibility, topology design for chiral photonic structures involving helically structured light has not been well investigated.
Specifically, the electromagnetism of topology-designed chiral optical antennas, acting as an interface between localized near field and propagating far field \cite{Hulst.2011rdw,Kawata.20151pk}, remains unexplored. In this study, using three-dimensional (3D) TO, we aimed to develop nanogap antennas with increased dissymmetry for improved CPL coupling.

{\it Topology design of nanogap antenna for CPL.}---TO began with the configuration illustrated in Fig.~1(a), involving a titanium dioxide (TiO$_2$) slab on a silicon dioxide (SiO$_2$) substrate.
\begin{figure}
\begin{center}
\includegraphics[width=8.6cm]{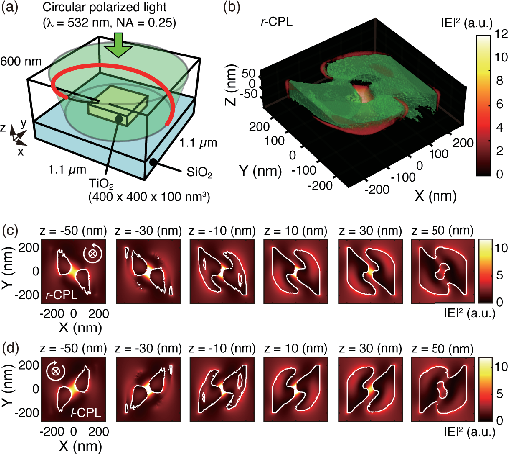}
\caption{
Topology-designed gap antenna for CPL. (a) Schematic of the boundary condition in TO.
(b) 3D rendering of the antenna structure (green) optimized for incident {\it r}-CPL. The electric field intensity distribution, calculated with the incident {\it r}-CPL, is overlaid. (c) {\it r}-CPL-optimized antenna structure (outlined in white) and the associated electric field intensity distribution at various $z$ positions. (d) Similar to (c) but optimized for incident {\it l}-CPL having reversed handedness.
}
\end{center}
\end{figure}%
A right-handed circularly polarized Gaussian beam with a wavelength of 532\,nm was directed onto the substrate via a 0.25 NA lens. The beam diameter was 982\,nm. 
We applied TO to maximize the power flux on an 8-nm$^3$ space at the center of the TiO$_2$ slab \cite{method}. A typical topology-optimized structure is represented in green in Fig.~1(b) (Supplementary Movies S1 and S2 for perspective views of the nanostructure). 
The topology-optimized shape exhibited a 3D vortex-like structure. Sections of the nanostructure are presented at 20-nm intervals in the $z$-direction in Fig.~1(c). 
(See Fig.~S1(a) for the pixel-by-pixel $z$-section.)
A twisted pattern with two bent arms was formed on the topmost surface ($z = 50$\,nm), resembling the gammadion shape previously proposed for plasmonic planar chiral metamaterials \cite{Kadodwala.2010,Kuwata-Gonokami.2011}. An aperture was created at the center of the top surface. Moving downward in the $z$-direction, the hole expands, dividing the structure into two parts at $z = 30$\,nm with a nanogap. The gap narrows to 12\,nm at $z = 0$\,nm, where the near-field intensity is maximum (Fig.~2(a)). The structure's base forms a diagonal wing-like shape at ($z = -50$\,nm).

Figure 1(d) shows the nanostructure obtained with the same configuration but for the opposite handedness of the incident circular polarization, namely left-handed CPL ({\it l}-CPL). 
The structure was mirror-reflected from that obtained for {\it r}-CPL (Fig.~1(c)). The observed mirror-asymmetry in these examples highlights the ability of TO to design chiral photonic nanostructures.

{\it Near-field properties of the gap mode: Field enhancement.}---To investigate the near-field properties of the topology-optimized structure, we performed finite-differential time-domain (FDTD) simulations on nanostructures optimized for {\it r}-CPL (Fig.~1(c)). Figures 2(a) and 2(b) compare the distributions of the near-field intensity excited by {\it r}-CPL (polarized co-circularly) and {\it l}-CPL (polarized cross-circularly) in the gap plane at $z = 0$\,nm.
\begin{figure}
\begin{center}
\includegraphics[width=8.6cm]{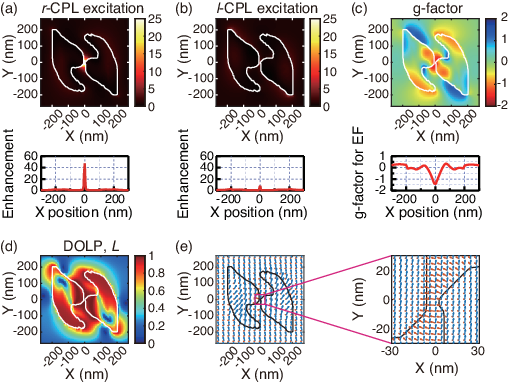}
\caption{
Near-field properties of the topology-designed antenna optimized for incident {\it r}-CPL. (a,b) Distributions of the intensity enhancement factors are shown for (a) {\it r}-CPL excitation (polarized co-circularly) and (b) {\it l}-CPL excitation (polarized cross-circularly), respectively, in the plane of the gap ($Z = 0$\,nm).
(c) Dissymmetry factor ($g$-factor) of the field intensity enhancement. The bottom panels display the sectional line profile along $Y = 0$\,nm. (d) Degree of linear polarization (DOLP), $L$. (e) Vector representation of the electric (red) and magnetic (blue) fields (left) and its magnified plot near the gap (right).
}
\end{center}
\end{figure}%
The color bars represent the enhancement factor, which is the ratio of the electric field intensity calculated with the antenna to that calculated without the antenna. The largest enhancement occurred at the gap position. TiO$_2$ acted as a dielectric at a designed wavelength of 532\,nm with a high refractive index ($n=2.512+0.011i$) \cite{method}, resulting in local field enhancement on the surface. The enhancement factors were 47.4 and 8.0 for co-circularly and cross-circularly polarized excitation, respectively, indicating that the gap antenna optimized for {\it r}-CPL preferentially coupled to the incident {\it r}-CPL with the same handedness. We compiled movies showing the optical near fields propagating on the surface of the gap antenna for both co-circular excitation ({\it r}-CPL) (Supplementary Movies S3--S5) and cross-circular excitation ({\it l}-CPL) (Supplementary Movies S6--S8) at different viewing angles. In the former scenario, the electric field propagated smoothly on the surface toward the nanogap. By contrast, in the latter scenario, the motion of the electric field was anti-phase and unrelated to the spiral shape of the structure.

We modified the dissymmetry factor \cite{Barron.2004}, originally refers to CD, to quantify the chiroptical properties of the topology-optimized antenna.
A very large dissymmetry factor of up to 1.13 was observed for plasmonic ramp-shaped nanostructures \cite{Wickramasinghe.2019}. 
Herein, the intensity of the excited gap modes was adopted
: $g_{gap} = 2(I_L-I_R)/(I_L+I_R)$, where $I_L$ and $I_R$ are the intensities of the gap modes excited by {\it l}-CPL and {\it r}-CPL, respectively. 
The $g$-factor ranges from $-2$ and $+2$, similar to the original one. The result is illustrated in Fig.~2(c). The $g$-factor peaked at $-1.42$ within the gap, demonstrating a giant chiral selectivity for the gap mode intensity against the incident CPL handedness.

{\it Polarization states.}---We investigated the impact of the gap antenna on incoming light polarization. To describe the near-field polarization, we calculated a dimensionless linear polarization parameter $L$ using generalized Stokes parameters \cite{method}. This parameter measures the degree of linear polarization (DOLP), where $L = 1$ indicates pure linear polarization and $L=0$ signifies circular polarization \cite{Sheppard.2014}. As shown in Fig.~2(d), the distribution of $L$ indicates that the near field within the gap is linearly polarized. The absence of the spin angular momentum (SAM) within the nanogap aligns with the reported result that a plasmonic planar dimer structure is incapable of receiving SAM from photons \cite{Sasaki.2023}. Figure 2(e) shows the electromagnetic field vectors around the gap. The electric field is oriented across the nanogap to reproduce the behavior of the gap-mode plasmon in a coupled plasmonic dimer system \cite{Abajo.2005}. The results demonstrate that the topology-designed chiral-gap antenna received CPL and transmitted it to the linearly polarized field within the gap as a near field.

{\it Coupled radiation of the point dipole with a chiral-gap antenna.}---The reciprocity principle in antenna theory, which states that the system is invariant under time reversal, leads us to hypothesize that the chiral-gap antenna radiates CPL coupled from a linearly polarized point dipole placed within the gap position of the antenna. This hypothesis was verified using an FDTD simulation. The configuration is shown in Fig.~3(a).
\begin{figure}
\begin{center}
\includegraphics[width=8.6cm]{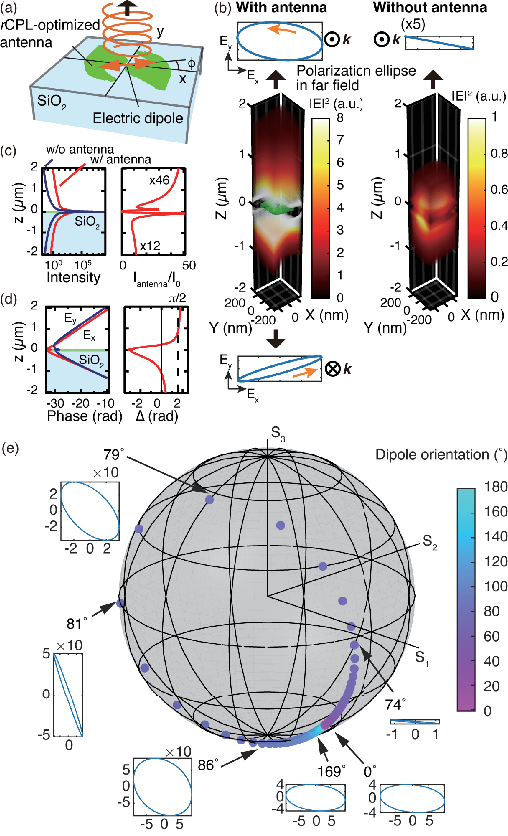}
\caption{
Coupled radiation of a point dipole with the topology-designed gap antenna optimized for {\it r}-CPL. (a) Schematic of the simulation model. 
(b) Electric field intensity of the dipole radiation with (left) and without (right) the gap antenna with the polarization ellipses (top and bottom panels) of the far field normal to the $X$-$Y$ plane.
Note that the substrate and the part of the field close to the source were intentionally rendered invisible for clarity.
(c) Field intensity with (red) and without (blue) the antenna along the $Z$ axis (left panel) along with the factor of the enhancement (right panel). 
The greenish part indicates the thickness of the antenna structure. (d) Phase angle of $E_x$ (red) and $E_y$ (blue) components along the $Z$ axis (left panel) and the phase difference (right panel). (e) Poincar\'e sphere representation of the far field polarization 
for different dipole orientation angles (color bar).
}
\end{center}
\end{figure}%
The gap antenna was the same as that optimized for {\it r}-CPL (Fig.~1(c)). A point dipole was placed in the middle of the gap. The in-plane orientation of the dipole was set parallel to the electric field vector assigned in Fig.~2(e) (the azimuth angle $\phi$ relative to the positive $x$-axis is 169.2$^\circ$). The entire structure was placed on a SiO$_2$ substrate to reproduce the same topology-designed configuration (Fig.~1(a)). The electric field intensities of the dipole radiation calculated with and without the gap antenna are shown in the left and right panels of Fig.~3(b), respectively.
The radiation intensified when the dipole was placed within the gap of the antenna structure, indicating that the gap antenna promoted the coupling of the near-field components of the electric dipole to the far field. The left panel of Fig.~3(c) illustrates the electric field intensity profile along the $z$-axis above and below the dipole located at $z = 0$\,nm. Without the antenna (dark blue curve), dipole radiation was distributed evenly in the forward and backward directions. By contrast, the antenna-coupled dipole radiation directed more energy in the forward direction (red curve). The asymmetric distribution in the radiation direction was apparent when the radiation intensity with the antenna was divided by that without the antenna, as shown in the right panel of Fig.~3(c). Forward scattering was approximately four times greater in intensity than backward scattering, indicating the antenna's capacity for directional control of radiation through its 3D shape. Previous demonstrations of the directional control of radiation have relied on principles such as those employing optical Yagi--Uda antenna \cite{Hofmann.2010} and metasurfaces \cite{Gaburro.2011}, which leverage the superposition (interference) of phase-shifted wave sources. Another mechanism is seen in dielectric Mie scattering, where the destructive superposition of electric and magnetic dipole scattering suppresses backward scattering, known as the Kerker condition \cite{Lukyanchuk.2012,Lukyanchuk.2016}. Unlike these mechanisms, the topology-designed 3D antenna realizes directional control of light scattering through the antenna shape design.

The left panel in Fig.~3(d) presents the phase angles of the $E_x$ and $E_y$ components along the $z$-axis. In the forward direction, the $E_x$ phase leads to $E_y$ by a nearly constant phase difference. The phase difference, shown in the right panel of Fig.~3(d), asymptotically approached $\pi/2$, suggesting that the resultant field vector rotated circularly in the counter-clockwise direction, i.e. left-handed rotation. As expected, the polarization ellipse in the far field, shown in the left panel of Fig.~3(b), traced an ellipsoidal trajectory. A point dipole is not purely represented as a linearly polarized source with a single-field vector orientation but is composed of near-field components with various field orientations, which partly explains the deviation from the perfect circle. For comparison, the right panel of Fig.~3(b) shows the same dipole radiation without the antenna, indicating that the far-field radiation was linearly polarized. The far-field polarization state can be decomposed into two orthogonal circular polarization states, $E_L$ and $E_R$, which were used to calculate the dissymmetry factor of the circular-polarized radiation, $g_{scat}=2(I_{scat,L}-I_{scat,R})/(I_{scat,L}+I_{scat,R})$, where $I_{scat,L}$ and $I_{scat,R}$ are the squared amplitudes of the corresponding components, $E_L$ and $E_R$, respectively. The calculated $g$-values were $1.46$ and $9.27\times10^{-7}$ with and without the gap antenna, respectively, demonstrating a significantly higher dissymmetry ($1.57\times10^6$-fold enhancement) in the coupled radiation of the linear dipole with the chiral-gap antenna.

In the backward direction (Figs.~3(b) and 3(d)), the phase difference between $E_x$ and $E_y$ components deviated from $\pi/2$, yielding a narrower ellipse with a smaller $g$-value of $-0.42$ in the far-field polarization state (Fig.~3(b)). The sign of the $g$-value in the backward direction was negative, indicating that backward scattering is polarized in {\it r}-CPL, which has the opposite handedness to forward scattering ({\it l}-CPL). Below, we will compare this finding with the structure obtained via counter-propagating CPL.

We also examined the effect of the dipole orientation on far-field radiation. Fig.~3(e) maps the far-field polarization states using Stokes parameters onto the Poincar\'e sphere for an orientation angle ranging from 0$^\circ$ to 180$^\circ$, along with representative polarization ellipses. The 180$^\circ$ rotation of the dipole orbited the polarization states around the Poincar\'e sphere, producing linear and circular polarization states that depend on the orientation angle. Notably, the polarization state approached the north pole, corresponding to {\it r}-CPL, most closely at an azimuthal angle of 79$^\circ$, when the dipole was oriented orthogonally to the electric field vector within the gap (Fig.~2(e)).

{\it Enhancing optical chirality with a topology-designed chiral-gap antenna.}---Optical chirality, expressed as $\chi = -\omega/(2c^2) \mathrm{Im}({\bf E^\ast \cdot H})$ for a monochromatic wave, is a physical quantity related to the enantioselective excitation of a chiral molecule when exposed to electromagnetic fields \cite{Cohen.2010,Cohen.2011}. Here, $\mathrm{\bf E}$ and $\mathrm{\bf H}$ are the complex, time-harmonic electric and magnetic fields in the frequency domain, respectively; $\omega$ is the angular frequency of the light; and $c$ is the speed of light in free space. To assess the impact of the structure on optical chirality inherent to CPL propagating in free space, we numerically calculated the optical chirality. The chirality values were normalized to that without the structure (i.e., the chirality value of {\it r}-CPL in free space, $\chi_{rCPL}=-\epsilon_0 \omega/(2c)E_0^2$, where $E_0$ is the incident electric field amplitude), as shown in Fig.~S4(a). The value reached at most 2.4 within the gap. The optical chirality was influenced by both the electric and magnetic field components and the mutual relation. We compiled a video of the electromagnetic field vectors in the time domain (Supplementary Movie S9). As shown in the video, the magnetic fields rotated, yielding nonparallel components against the linearly polarized electric fields, with their orientation fixed across the gap. Moreover, within the gap, the phase angle between $i{\bf E}$ and ${\bf H}$ significantly deviated from that of {\it r}-CPL in free space having $\cos(\delta_{i{\bf E},{\bf H}})=1$ (Fig.~S4(b)). Based on these analyses, we anticipate that chirality will be further enhanced by making the $i{\bf E}$ and ${\bf H}$ components aligned in phase with parallel orientation.

To achieve the parallel orientation of the electric and magnetic field vectors, we extended the excitation scheme to incorporate counter-propagating {\it r}-CPL, as shown in Fig.~4(a).
\begin{figure}
\begin{center}
\includegraphics[width=8.6cm]{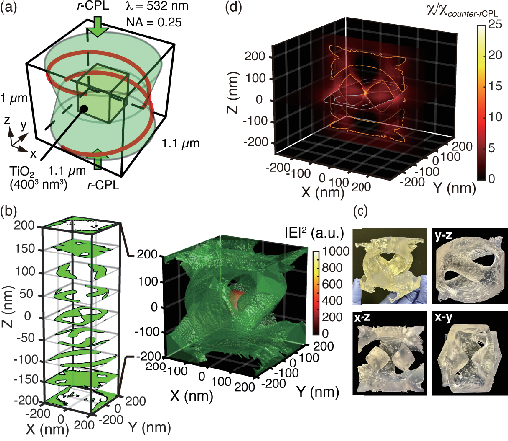}
\caption{
Topology-designed gap antenna for counter-propagating {\it r}-CPL. (a) Schematic of the boundary condition. (b) 3D-rendered model of the topology-designed antenna (right) and $Z$-sliced sections 
(left). The electric field intensity distribution when excited with counter-propagating {\it r}-CPL is overlaid.
(c) 3D-printed mockup of the chiral gap antenna at a scale of 140,000/1.
(d) Optical chirality density, $\chi$, normalized to that without the structure, displayed at planes $Z = 0$\,nm and $Y = 0$\,nm. The excitation is counter-propagating {\it r}-CPL.
} 
\end{center}
\end{figure}%
The thickness of the TiO$_2$ layer was quadrupled to accommodate excitation from both sides. The resultant structure is depicted in Fig.~4(b) as a 3D-rendered model. Figure 4(c) displays images of a scaled mock-up at a ratio of 140,000:1 created using a commercial 3D printer, demonstrating its capability to stand freely as a structure. A complete set of $z$-sections is shown in Fig.~S5. A perspective view of the nanostructure is presented in Supplementary Movies S10 and S11. An 8-nm-wide nanogap is created at the center of the plane $z = 0$\,nm. The geometry above the gap plane ($z = 0$\,nm) exhibited a helical winding structure, forming a mirror image of its counterpart beneath the gap plane. As shown in the right panel of Fig.~4(b), the electric field was localized and enhanced at the center of the gap. The enhancement factor of the electric field intensity peaked at 1,250 for the field intensity of the same counter-propagating {\it r}-CPL that did not include the structure. Supplementary Movies S12 and S13 feature optical near fields propagating on the gap antenna. The former involves {\it r}-CPL (co-circular) excitation, whereas the latter involves cross-circular {\it l}-CPL excitation.
The enhancement factors of the electric field intensity calculated for {\it r}-CPL and {\it l}-CPL excitations, $g_\mathrm{gap}$, and polarization state near the gap are summarized in Fig.~S6. The extremum of the value $g_\mathrm{gap}$ reached as low as $-1.70$ (Fig.~S6(c)) within the gap, indicating an even greater dissymmetry, compared to the structure optimized for a single {\it r}-CPL (Fig.~2(c)). As shown, in Fig.~S6(e) and Movie S14, the electric and magnetic field vectors were oriented in parallel within the gap, a preferable condition for rnhancing optical chirality. The distribution of optical chirality, obtained for counter-propagating {\it r}-CPL excitation,
is displayed in Fig.~4(d). The enhancement factor reached a maximum of 24.5 in the middle of the gap, more than 10-times greater than the case optimized with a single {\it r}-CPL (2.4-fold enhancement). The optical chirality of a counter-propagating {\it r}-CPL in free space has a constant value throughout space and is double that of a single CPL with the same amplitude \cite{Cohen.2010}. Consequently, the structure provides approximately a 50-fold increase in the optical chirality density within the nanogap compared with that of CPL propagating in free space. $i{\bf E}$ and ${\bf H}$ also remained in phase within the gap ($\cos(\delta_{i{\bf E},{\bf H}})=0.98$), as shown in Fig.~S7(b).

Optical chirality is a conservative quantity, akin to a Poynting vector \cite{Lipkin.1964}. The corresponding continuity equation is given as $d\chi/dt + \nabla \cdot {\bf F} = 0$, where ${\bf F}=(\omega/4)\mathrm{Im}(\epsilon_0{\bf E^\ast \times E} + \mu_0{\bf H^\ast \times H})$ is the optical chirality flux and is proportionally related to the SAM density by a factor of $\omega^2$ \cite{method,Nori.2017,Shimura.2020,method}. In Fig.~5, we visualize the SAM density ${\bf s}$
together with $\nabla \cdot {\bf s}$.
\begin{figure}
\begin{center}
\includegraphics[width=8.6cm]{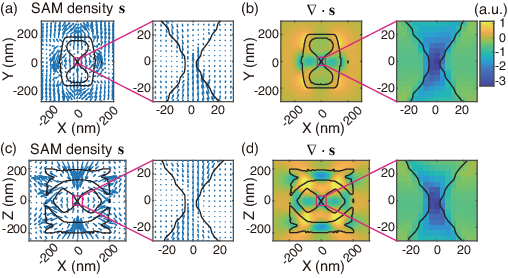}
\caption{
(a) SAM density vector in the $X$-$Y$ plane at $Z = 0$\,nm (left) and the magnified view (right) and (b) the corresponding divergence of SAM. 
(c,d) Same as (a,d) but the $X$-$Z$ plane at $Y = 0$\,nm, respectively.} 
\end{center}
\end{figure}%
As shown in the vector plot, the SAM density converged inward in the presence of the gap antenna. With opposite signs for the counter-propagating {\it r}-CPLs, the flow of the SAM canceled out at the gap position, creating a distinct negative divergence (sink) of the SAM, as displayed in Figs.~5(b) and 5(d). This led to an increase in the optical chirality.

Figure S8 shows the coupled radiation of a linear dipole embedded in the middle of the gap.
Unlike the results obtained for the chiral-gap antenna optimized with single  {\it r}-CPL (Fig.~3(b)), the far-field polarization states were  {\it l}-CPL for both the forward and backward scattering directions, with an even higher dissymmetry value of $g_{scat} = 1.6$.

{\it Discussion.}---We showcased enhanced optical chirality in gap mode configurations, facilitating the combination of confined mode volumes for field and optical chirality enhancement. This augments surface-enhanced Raman optical activity spectroscopy efficiency \cite{Fang.2018, Goda.2021}. The demonstrated CPL radiation from the point dipole coupled with the chiral gap antenna offers potential for circularly polarized luminescence devices. While CPL emission is evident in specific organic molecules \cite{Riehl.1977}  and explored for emerging tech applications, its luminescence $g$-factoris usually small, approximately $10^{-3}$ \cite{Nakashima.2007}.
Our gap antenna presents an innovative approach for CPL-emitting devices by converting the polarization from linear emitters \cite{Kuwata-Gonokami.2011}.

The initial work introducing optical chirality examined a system comprising similar counter-propagating CPL but without the presence of structures \cite{Cohen.2010}. 
This work represents a straightforward extension by incorporating chiral nanogap structures and the associated near-field interactions into the illumination scheme of counter-propagating CPL, demonstrating the excellent performance in enhancing both the local optical chirality density and the near-field intensity at the nanoscale.
The proposed design approach utilizing TO holds significant potential for realizing advanced nanophotonic structures for future chiral applications, as well as for nanophotonic control of more intricate structured light including optical vortex \cite{Woerdman.1992}, skyrmion \cite{Bartal.2018jhm,Rosales-Guzman.2022}, and hopfions \cite{Zayats.2023}.


Topology-designed structures, with their intricate 3D geometries, pose a challenge for fabrication using traditional micro/nanofabrication methods like photolithography and electron beam lithography, which are largely planar \cite{Munchmeyer.1986}. Recent progress in laser 3D fabrication allows for more flexible creation of 3D shapes \cite{Chen.2022,Duan.2022}, with a growing trend of incorporating inorganic materials \cite{Duan.2022} like TiO$_2$ and ZrO$_2$ \cite{Fujita.2020}.


\phantom{\cite{Sigmund.2021,Fan.2012tug,Palik.1997}}

\begin{acknowledgments}
We thank the use of the AI-compatible advanced computer system at the Information Initiative Center, Hokkaido University, Sapporo, Japan, under the joint research program.
This study is supported by a Grant-in-Aid for Transformative Research Areas, ``Evolution of Chiral Materials Science using Helical Light Fields'' 
(Grants No. JP22H05131, No. JP22H05137) from JSPS KAKENHI, Japan. 
\end{acknowledgments}


%

\end{document}